\documentclass[usenatbib]{mn2e}
\usepackage[pdftex]{graphicx}
\usepackage{indentfirst}
\usepackage{amssymb}
\usepackage{amsmath}
\usepackage{verbatim}

\begin{document}
\title{On linear dust-gas streaming instabilities in protoplanetary discs}
\author[Jacquet et al.]{Emmanuel Jacquet,$^1$ Steven Balbus,$^2$ Henrik Latter$^3$\\
$^1$LMCM, Mus\'{e}um National d'Histoire Naturelle, 57 rue Cuvier, 75005 Paris, France\\
$^2$LRA, Ecole Normale Sup\'{e}rieure, 45 rue d'Ulm, 75005 Paris, France\\
$^3$AFD Group, DAMTP, University of Cambridge, CMS Wilberforce Rd, Cambridge UK CB3 0WA
}

\maketitle

\begin{abstract}
We revisit, via a very simplified set of equations, a linear streaming
instability (technically an overstability), which is present in, and
potentially important for, dusty protoplanetary disks
\citep{YoudinGoodman2005}. The goal is a
better understanding of the physical origin of such instabilities, which
are notoriously subtle.  
Rotational dynamics seem to be essential to this
type of instability, which cannot be captured by one-dimensional Cartesian models. 
 Dust `pileups' in moving pressure maxima are an important
triggering mechanism of the instability, and drag feedback of dust upon the gas  allows these maxima to be strengthened. Coriolis forces and the background drift counteract the effects of the pressure force.

\end{abstract}

\section{Introduction}

The formation of planets via dust accretion starting with
micron-sized grains and ending with $10^4-10^5$-km-sized bodies in
a protoplanetary disc is a multiphase process (see e.g. the review by
\citet{ChiangYoudin2010}, henceforth CY10). Grains smaller than roughly
a centimetre coagulate via sticking \citep[e.g.][]{Dominiketal2007},
while for the largest sizes, gravity of the planetary bodies plays a
dominant role \citep[e.g.][]{Ida2010}.

Between these two limits, there is a gap that neither mechanism
seems able to leapfrog. Beyond a size of 1 cm - 1 m, collisions
occur at velocities of order $1-10\:\mathrm{m\:s^{-1}}$.  These
lead either to fragmentation or to bouncing \citep{Zsometal2010},
rather than to systematic aggregation. Worse yet, particles of this
size will drift sunward at the dramatic rate of 1 AU per century
\citep{Weidenschilling1977}.  This is because solids tend to orbit the
Sun at the full Keplerian speed, while the gas moves slightly slower
because of the support from an outward directed pressure force.
Thus, the solids experience a headwind that removes angular momentum
via gas drag. This effect is the cause of the ``metre-size barrier''.

Self-gravity of the dust could in principle promote accretion past this barrier
provided some process enhances the dust density.
A classic scenario for this is one in which the vertical component of
the Sun's gravity causes sedimentation to the disc midplane and increases
the dust density \citep{GoldreichWard1973}. Although \citet{Weidenschilling1980} had suggested that Kelvin-Helmholtz instabilities could prevent the Roche density from being reached, recent studies indicate that this is not an issue for discs a few times more massive than the minimum mass solar nebula, or with supersolar metallicities \citep[e.g.][]{Chiang2008,Leeetal2010a,Leeetal2010b}. Moreover, \citet{Youdin2011} showed that inclusion of gas drag could alleviate the need to actually reach the Roche density. However, these mechanisms assume a low level of global turbulence. While the magnetorotational instability (\citealt{BalbusHawley1998}) may be suppressed in the so-called dead zone \citep{Gammie1996}, fluctuations arising from forced density waves or other instabilities could still be present \citep{FlemingStone2003,LesurPapaloizou2010,Latteretal2010}. Concentrating the dust is problematic.

However, interesting effects already occur as soon as the dust-to-gas ratio is of order unity, say around the midplane, or as a result of turbulent concentration \citep[e.g.][]{Cuzzietal2003}. This is because the dynamical back-reaction of the dust on the gas is then important, while self-gravity itself may still be negligible. In
fact, \citet{YoudinGoodman2005} (hereafter YG05) discovered that the
interpenetration of dust and gas in a Keplerian disc was linearly
overstable: a process that they named the \textit{streaming
instability}, a convention we shall henceforth adopt. The dust particles can behave
collectively via their
interaction with the gas, leading to disordered flows that facilitate grain clumping---as witnessed by the numerical simulations of \citet{YoudinJohansen2007} (hereafter YJ07)
and \citet{JohansenYoudin2007}. \citet{BaiStone2010a} confirmed that
in the nonlinear regime, dust densities up to three orders of magnitude
above the background could be attained, which \citet{Johansenetal2007}
argued would suffice to instigate gravitational clustering. As relatively large (10 cm - 1 m) particles are needed for significant clumping \citep{BaiStone2010a}, it remains to be demonstrated whether solids can have grown this large beforehand \citep{Leeetal2010b}.

Although much attention has been devoted to the streaming instability
in the past years, its physical interpretation in the \textit{linear}
phase remains somewhat obscure.  For example, as YJ07 point out, the idea
that radial drift slows down in overdense regions and leads to a sort of
`traffic jam', does not explain the onset of growth in the linear phase
---such reasoning leads, in fact, to stable wave
propagation. Such traffic jams, however,
appear to be relevant for the nonlinear phase (JY07).  JY07 showed that
drafting, analogous to pelotons in bicycle races, also was not necessary
for the initial generation of particle overdensities, although it could enhance
growth if present.  CY10
discuss a one-dimensional vertically integrated toy model drawn from
\citet{GoodmanPindor2000}, in which a (single) fluid is subject to a
``collective drag'' acceleration in the sense that a drag coefficient
depending upon the fluid density emerges from collective behavior.
This system is found to be overstable.  It is unclear, however, in what
sense the streaming instability conforms to this notion of collective
drag since it is local in nature: the equations are not vertically
integrated and the drag acceleration does not depend {\it ab initio}
on the density of the fluid acted upon.

We are therefore motivated to reconsider the origins of dust clumping.
In this paper, we revisit the linear streaming instability. In \S 2,
we review the basic equations leading to the YG05 results 
and propose, in an appropriate limit, a markedly reduced system in which
the instability arises.  We make the point in \S 3  that rotational
dynamics are essential to understand the streaming instability. In \S 4,
we propose an interpretation of the streaming instability. In \S 5,
we summarise our conclusions.

\section{Streaming instability and reduced systems}

In this section, by way of establishing notation (largely that of YG05),
we briefly review the fundamental equations along with some of their
more important results. We then propose a significant reduction of the
system of equations that involves two successive levels of approximation,
yet still produces the streaming instability.

\subsection{Fundamental two-fluid equations and local stability analysis}

Let us model the dust as a collection of identical, indestructible spheres, of
mass density $\rho_p$ and velocity $\mathbf{V}_p$. The gas density and
velocity, on the other hand, we denote by $\rho_g$ and $\mathbf{V}_g$,
respectively. Their evolution equations are (see Appendix A for justification
of
the dust equations):
\begin{equation}
\frac{\partial \rho_p}{\partial t} + \nabla\cdot(\rho_p\mathbf{V}_p)=0
\label{continuity-dust}
\end{equation}
\begin{equation}
\frac{\partial \rho_g}{\partial t} + \nabla\cdot(\rho_g\mathbf{V}_g)=0
\label{continuity-gas}
\end{equation}
\begin{equation}
\mathcal{D}_p\mathbf{V}_p=-\Omega^2 \mathbf{R} -
\frac{\mathbf{V}_p-\mathbf{V}_g}{t_{\mathrm{stop}}}
\label{Euler-dust}
\end{equation}
\begin{equation}
\mathcal{D}_g\mathbf{V}_g=-\Omega^2 \mathbf{R}
+\frac{\rho_p}{\rho_g}\frac{\mathbf{V}_p-\mathbf{V}_g}{t_{\mathrm{stop}}}-\frac{\nabla
P}{\rho_g},
\label{Euler-gas}
\end{equation}
where $\Omega$ is the Keplerian angular velocity,
$\mathcal{D}_{p,g}\equiv\partial /\partial t+\mathbf{V}_{p,g}\cdot\nabla$
the particle/gas Lagrangian derivatives, and $\mathbf{R}$ the cylindrical
vector radius.   We ignore the vertical component of the solar gravity, as
well as self-gravity.

The particle stopping time $t_{\rm stop}$ depends on the size and
velocity regime \citep[e.g.][]{Weidenschilling1977}. For example, for particles
that are both small compared to the gas mean-free-path and drifting
subsonically relative to the gas (as appropriate, e.g., for chondrule-sized
bodies at a few AUs in a Minimum Mass Solar Nebula), Epstein's law
\citep{Epstein1924} applies:
\begin{equation}
t_{\mathrm{stop}}=\frac{\rho_s a}{\rho_g v_T},
\end{equation}
with $v_T=\sqrt{8k_BT/\pi m}$ (with $m$ the molecular mass and $T$ the
temperature), roughly the sound speed, $\rho_s$ the \textit{internal} density
of the grains (not to be confused with $\rho_p$) and $a$ the grain radius.
Regardless of the relevant drag law, one defines a dimensionless stopping time
measuring the coupling of dust to dynamical disturbances:
\begin{equation}
\tau_s\equiv\Omega t_{\mathrm{stop}},
\end{equation}
By virtue of Newton's third law, an all-important feedback term
of the dust on the gas appears in the Euler equation for the gas
(\ref{Euler-gas}).
  The system of equations is closed by assuming
gas incompressibility (the Boussinesq approximation), as in
YG05.

  It is convenient to express equations
(\ref{continuity-dust})-(\ref{Euler-gas}) in terms of centre-of-mass velocity
$\mathbf{V}\equiv (\rho_p\mathbf{V}_p+\rho_g\mathbf{V}_g)/\rho$ (with
$\rho\equiv\rho_p+\rho_g$ the total density) and the relative dust-to-gas drift
$\Delta\mathbf{V}\equiv\mathbf{V}_p-\mathbf{V}_g$ (YG05):
\begin{equation}
\frac{\partial \rho}{\partial t}+\nabla\cdot(\rho\mathbf{V})=0
\label{continuity-COM}
\end{equation}
\begin{equation}
\nabla\cdot(\mathbf{V}-\frac{\rho_p}{\rho}\Delta\mathbf{V})=0
\label{incompressibility}
\end{equation}
\begin{equation}
\frac{\partial \mathbf{V}}{\partial
t}+\mathbf{V}\cdot\nabla\mathbf{V}+\mathbf{F}(\Delta\mathbf{V}^2)=-\Omega^2\mathbf{R}-\frac{\nabla
P}{\rho}
\label{Euler-COM}
\end{equation}
\begin{equation}
\frac{\partial\Delta\mathbf{V}}{\partial
t}+\mathbf{V}\cdot\nabla(\Delta\mathbf{V})+(\Delta\mathbf{V}\cdot\nabla)\mathbf{V}+\mathbf{G}(\Delta\mathbf{V}^2)=-\frac{\rho}{\rho_g}\frac{\Delta\mathbf{V}}{t_{\mathrm{stop}}}+\frac{\nabla
P}{\rho_g},
\label{Euler-DeltaV}
\end{equation}
with:
\begin{equation}
\mathbf{F}(\Delta\mathbf{V}^2)\equiv\frac{1}{\rho}\nabla\cdot\left(\frac{\rho_g\rho_p}{\rho}\Delta\mathbf{V}\Delta\mathbf{V}\right)
\end{equation}
\begin{equation}
\mathbf{G}(\Delta\mathbf{V}^2)\equiv\frac{\rho_g}{\rho}\Delta\mathbf{V}\cdot\nabla\left(\frac{\rho_g}{\rho}\Delta\mathbf{V}\right)-\frac{\rho_p}{\rho}\Delta\mathbf{V}\cdot\nabla\left(\frac{\rho_p}{\rho}\Delta\mathbf{V}\right).
\end{equation}
 We shall now work in the so-called shearing-sheet approximation: We neglect
all curvature terms and disc-scale gradients since the lengthscales of interest
are much smaller than the heliocentric distance, or even the pressure scale
height. We also adopt Cartesian coordinates in a frame corotating with the
Keplerian flow at a fixed heliocentric distance $R_0$, at angular velocity
$\Omega$. We denote by $x$, $y$ and $z$ coordinates in the radial, azimuthal,
and vertical (that is, perpendicular to the midplane) directions, respectively,
and $\mathbf{e}_x$, $\mathbf{e}_y$ and $\mathbf{e}_z$ the corresponding unit
vectors.

  Using this model, \citet{Nakagawaetal1986} computed the
equilibrium solution to equations \eqref{continuity-COM}-\eqref{Euler-DeltaV},
\begin{equation}
\mathbf{V}=(-\frac{g_e}{2\Omega}-\frac{3}{2}\Omega x) \mathbf{e_y}
\label{NSH-V}
\end{equation}
\begin{equation}
\Delta\mathbf{V}=-\frac{g_et_{stop}}{1+(f_g\tau_s)^2}\mathbf{e_x}+\frac{f_gg_e\Omega
t_{stop}^2}{2(1+(f_g\tau_s)^2)}\mathbf{e_y},
\label{NSH-DeltaV}
\end{equation}
where $f_{g,p}\equiv\rho_{g,p}/\rho$ the mass fraction of gas and particles (of
uniform densities in this equilibrium flow), and
\begin{equation}
 g_e\equiv-\frac{1}{\rho}\frac{dP}{dR}
\end{equation}
is the pressure-induced acceleration on the gas+dust fluid. It is the
free energy stored in this gradient that the streaming instability can access (YG05).
This gradient, consequently,
 introduces a lengthscale which helps determine the
characteristic scale of the instability. We denote this scale by $L$
and quantify it through:
\begin{equation} \label{LLL}
L= \frac{g_e}{\Omega^2} \sim \left(\frac{H}{R}\right)H,
\end{equation}
where $H$ is the gas disc semi-thickness. It follows that $L$ is a
small fraction of $H$.

  We now decompose the variables in equilibrium value and perturbation as
follows:
\begin{equation}
\mathbf{V}=\mathbf{V}_0+\mathbf{v}
\end{equation}
\begin{equation}
\Delta\mathbf{V}=\Delta\mathbf{V}_0+\Delta\mathbf{v}
\end{equation}
\begin{equation}
\rho=\rho_0(1+\delta)
\end{equation}
\begin{equation}
P=P_0(x)+\rho_0h,
\end{equation}
where the zero subscripts refer to the unperturbed background. Coordinates of
the velocity perturbations are given by:
$\mathbf{v}=u\mathbf{e}_x+v\mathbf{e}_y+w\mathbf{e}_z$,
$\Delta\mathbf{v}=\Delta u\mathbf{e}_x+\Delta v\mathbf{e}_y+\Delta
w\mathbf{e}_z$. We then decompose the perturbation into
 axisymmetric Fourier modes
$\propto \exp{[i(k_xx+k_zz-\omega t)]}$.
Here, $k_x$ and $k_z$ are real wavenumbers and $\omega\equiv\omega_{\Re}+is$ is a
complex frequency, with $\omega_{\Re}$ a wave frequency and $s$ a growth rate
(if $s<0$, the perturbation is damped).

  YG05 studied the resulting sixth-order system numerically and found unconditional instability, regardless of the value of the dimensionless parameters $\tau_s$
and $f_p$, with a subdynamical growth rate scaling like $\Omega\tau_s$ in the
tight-coupling limit. For a given $\tau_s$ and $f_p$, growth was maximised in
the $k_x$-$k_z$ plane in a long-wavelength ridge (with $k_z\sim k_x^2
t_{\mathrm{stop}}/(\Omega f_g)$) and a short-wavelength branch with $k_z \gg
k_x$ ($k_x$ having a preferred value). Growth was strongly suppressed in the
former for dust-to-gas ratio near unity ($f_p\sim f_g\sim 0.5$). YG05 found
that the wave speed $\omega_\Re/k_x$ was bounded by $U_{\mathrm{sum}}$ defined
as the sum of the background radial velocities of the dust and the gas,
respectively, and growth rates (generally smaller than
$|k_x|U_{\mathrm{sum}}/2$) were maximised when the said wave speed was about
$U_{\rm sum}/2$. 

\subsection{Terminal velocity approximation}

To gain physical insight on the streaming instability, we need some
simplifications to make the problem tractable analytically.  We shall
adopt the ``terminal velocity approximation'' (YG05), that is take the
left-hand-side of equation (\ref{Euler-DeltaV}) to be zero, such that
the relative drift is given by:
\begin{equation}
\Delta\mathbf{V}=\frac{\nabla P}{\rho}t_{\mathrm{stop}}.
\label{TVA}
\end{equation}
This approximation holds provided (i) the perturbation and dynamical
timescales ($\omega^{-1}$ and $\Omega^{-1}$) are longer than
$f_gt_{\mathrm{stop}}=\rho_sa/(\rho v_T)$ (where the equality holds in
the Epstein drag regime) and (ii) the lengthscale $1/k$ of variation
is longer than the ``stopping length'' $f_gg_et_{\mathrm{stop}}^2=f_g\tau_s^2L$
(see Appendix B for justification).

We will additionally neglect $\mathbf{F}$ and $\mathbf{G}$ because of
the smallness of $\Delta\mathbf{V}$, an approximation YG05 find to be
safe for $\omega\ll\Omega$.

  The linearised system of equations is then:
\begin{equation}
-i\omega u -2\Omega v +g_e\delta +ik_xh =0
\label{quartic system Euler x}
\end{equation}
\begin{equation}
\frac{\kappa^2}{2\Omega}u -i\omega v=0
\label{quartic system Euler y}
\end{equation}
\begin{equation}
-\omega w +k_zh =0
\label{quartic system Euler z}
\end{equation}
\begin{equation}
ik_xu +ik_zw -i\omega\delta=0
\label{quartic system COM continuity}
\end{equation}
\begin{equation}
ik_xu+ik_zw-ik_xg_e(f_p-f_g)t_{\mathrm{stop}}\delta +f_pk^2t_{\mathrm{stop}}h
=0
\label{quartic system incompressibility}
\end{equation}
where zero subscripts have been dropped, $k^2=k_x^2+k_z^2$,
and we have introduced the epicyclic frequency $\kappa$, given by
$\kappa^2\equiv R^{-3}d(R^4\Omega^2)/dR$.
In a Keplerian disc, $\kappa=\Omega$, but one could also imagine studying
streaming instabilities
in any rotating fluid,
possibly even by direct experimentation. Equations \eqref{quartic system Euler x}-\eqref{quartic system Euler z} are derived from perturbation of the $x$, $y$ and $z$ components
of equation (\ref{Euler-COM}).   Equations \eqref{quartic system COM continuity} and \eqref{quartic system incompressibility} stem from
perturbing the continuity equations
(\ref{continuity-COM})
and
(\ref{incompressibility})
respectively.  (In equation [\ref{quartic system incompressibility}] we also make use of equation [\ref{TVA}].)
Notice that the important term proportional to ($f_p-f_g$) in equation \eqref{quartic system incompressibility} derives from a
sort of buoyancy force that arises in the two fluid systems, with an effective
density of $\rho^2/\rho_p$.  When $f_p=f_g$, this effective density is
stationary with respect to perturbations in $\rho_p$.

One can straightforwardly obtain a dispersion relation by setting the
determinant of the above 5$\times$5 system to zero\footnote{We note
that the first term of the quadratic coefficient differs from that
of equation (39) of YG05.  While we believe that equation (\ref{22}) of the
current
paper is correct (equation (\ref{growth-TVA}) corrects their equation (44)
accordingly), the YG05 error appears to be typographical and in no event does
it affect
their exact results.
Also, YG05 drop the quartic term in their equation (39); this is however
consistent with $\omega t_{\mathrm{stop}}\ll 1$.}:
\begin{eqnarray}\label{22}
-if_pt_{\mathrm{stop}}\omega^4 + \omega^3 +
(if_p\kappa^2+k_xg_ef_g)t_{\mathrm{stop}}\omega^2-\left(\kappa\frac{k_z}{k}\right)^2\omega\nonumber\\
 + k_x\left(\kappa\frac{k_z}{k}\right)^2g_et_{\mathrm{stop}}(f_p-f_g)=0
\label{quartic}
\end{eqnarray}
Series solutions in $\tau_s$ carried out by YG05 have shown that the roots fall
in three branches, two epicycles giving rise to damping, 
 and a secular mode, with:
\begin{equation}
\omega_\Re=k_x(f_p-f_g)g_et_{\rm stop}+o(t_{\rm stop})
\label{omegaR}
\end{equation}
\begin{equation}
s=if_pt_{\mathrm{stop}}^3\left((f_p-f_g)g_e\frac{kk_x}{k_z}\right)^2+o(t_{\mathrm{stop}}^3),
\label{growth-TVA}
\end{equation}
to leading order in the stopping time\footnote{It must be cautioned that
the terminal velocity approximation discards terms of third and higher
order in $t_{\mathrm{stop}}$ that might contribute to the leading order
expansion of the \textit{growth rate} and hence could in principle affect
its sign (which the YG05 calculations show not to be the case here). Since we
are here interested in understanding which ingredients
give rise to instability in a model system, rather than studying its exact
properties
(as in YG05), we content ourselves with this heuristic approach.}.  (The
notation
$o(X)$ indicates that the ratio $o(X)/X$ tends to zero as $X\rightarrow 0$). It
can be seen that feedback is essential to the instability, as ignoring it would
amount to setting $f_p=0$ everywhere. Also, the necessity of the background drift as measured by $g_e$ is evident. More rigorously, it may be seen that for $g_e=0$, the dispersion relation becomes, if we disregard the $\omega=0$ mode:
\begin{equation}
f_p\kappa t_{\rm stop}X^3-X^2+f_p\kappa t_{\rm stop}X-\left(\frac{k_z}{k}\right)^2=0
\end{equation}
with $X\equiv i\omega/\kappa$. One can show that, for sufficiently small $\kappa t_{\rm stop}$, this cubic has three real positive roots (in terms of $X$), corresponding to $\omega$ being a purely imaginary number of negative imaginary part
.

\subsection{The secular mode in a reduced system}

It is possible to simplify the system further while retaining the leading-order
physics of the instability.
First, we note that, in general, $\omega\ll\Omega$, as may be judged from the exact
results of YG05 or equation (\ref{omegaR}), for wavelengths that are not
significantly shorter than the radial pressure scale $L$.
 Equation \eqref{quartic system Euler y} gives us
$$u=(2i\Omega\omega/\kappa^2)v\sim (\omega/\Omega)v.$$ It is then easy to show
 that in equation \eqref{quartic system Euler x} the first term
(the acceleration term $-i\omega u$) can be neglected.
It follows that, to leading order, the
$x$ and $y$ equations of motion of the centre of mass relax to a form of
\emph{geostrophic balance}, i.e. Coriolis forces effectively cancel out the pressure
gradient and the buoyancy force. This is important, as it means that pressure
perturbations do not effectively drive centre-of-mass motions in the orbital plane. So this crucial
stabilising tendency is consigned to a subdominant role. It also means
that the centre of mass executes modified epicycles in the orbital
plane, to
leading order.

Second, we neglect the $u$ perturbation in the two continuity equations
\eqref{quartic system COM continuity} and \eqref{quartic system incompressibility} which we are permitted to do if we restrict
ourselves to long radial wavelengths. From equations
\eqref{quartic system Euler x}-\eqref{quartic system Euler y} we can obtain the following scalings
$$  u \sim \frac{\omega}{\Omega}\, v \sim 
\mathrm{max}(k_x\frac{\omega}{\Omega^2}\,h, \omega\frac{g_e}{\Omega^2}\,\delta).
$$
Combining these with \eqref{omegaR} and equation
\eqref{quartic system Euler z}, we see that $k_x u$ is subdominant here if\footnote{The second condition results from requiring $k_x(k_x\omega h/\Omega^2)\ll k_zw$ with $h$ eliminated from equation \eqref{quartic system Euler z}.} 
\begin{align}
k_x\, L \ll 1, \qquad \text{and}\qquad \left(\frac{k_x}{k_z}\right)\left(\frac{\omega}{\Omega}\right)\ll 1,
\label{condition}
\end{align} 
which can be satisfied if $k_x H \sim 1$ (see equation \eqref{LLL})
and if $k_x/k_z\lesssim 1$.

These approximations give us a simplified set of dynamical equations:
 the equations for the vertical centre
 of mass velocity, particle density conservation, and the
 incompressibility of the gas, in which drag effects enter
 implicitly. These are now

\begin{equation}
-i\omega w +ik_zh =0
\label{quadratic system Euler z}
\end{equation}
\begin{equation}
ik_zw -i\omega\delta=0
\label{quadratic system COM continuity}
\end{equation}
\begin{equation}
ik_zw -ik_xg_e(f_p-f_g)t_{\mathrm{stop}}\delta +f_pk^2t_{\mathrm{stop}}h =0
\label{quadratic system incompressibility}
\end{equation}
yielding the following quadratic dispersion relation
:
\begin{equation}
-if_pt_{\mathrm{stop}}\left(\frac{k}{k_z}\omega\right)^2 + \omega +
k_xg_et_{\mathrm{stop}}(f_g-f_p)=0,
\label{quadratic-reduced}
\end{equation}
from which we recover the secular mode and its growth rate as
given in equation
\eqref{growth-TVA}! The corresponding eigenvector reads:
\begin{equation}
\left[\begin{array}{r}
w \\ \delta \\ h
\end{array}\right]
=\delta
\left[\begin{array}{r}
\omega/k_z \\ 1 \\ \left(\omega/k_z\right)^2
\end{array}\right]
\label{eigenvector}
\end{equation}
 The second root corresponds to a damping
$s=-\left(k/k_z\right)^2/(f_pt_{\mathrm{stop}})$ but violates the condition $\omega t_{\rm stop}\ll 1$ of validity of the terminal velocity approximation and will thus not be discussed
further here.

  It is also possible, if less rigorous, to see this reduction directly by inspection of the quartic dispersion relation (\ref{quartic}). If we have $k_xg_et_{\mathrm{stop}}^2(k_x/k_z)^2\ll 1$ and $k_xg_e/\kappa^2\ll 1$ (which imply the conditions of \eqref{condition}), 
one is allowed to modify the quartic term as $-if_pt_{\mathrm{stop}}\left(k/k_z\right)^2\omega^4$ since $if_p\left(k_x/k_z\right)^2t_{\mathrm{stop}}\omega^4 \ll \omega^3$ and the quadratic one as $(if_p\kappa^2+k_xg_e(f_g-f_p))t_{\mathrm{stop}}\omega^2$ since $k_xg_ef_p\ll if_p\kappa^2$. Under these conditions, the quartic may be factored as:
\begin{eqnarray}
\left(\omega^2-\left(\kappa\frac{k_z}{k}\right)^2\right)\left(-if_pt_{\rm stop}\left(\frac{k}{k_z}\right)^2\omega^2+\omega+k_xg_et_{\mathrm{stop}}(f_g-f_p)\right)\nonumber
\end{eqnarray}
such that we retrieve the quadratic dispersion relation of the reduced system.

Hence, it appears that the mathematical essence of the instability can be
 isolated if we make the following set of assumptions: (a) the
 terminal velocity approximation, in which
 stopping time is short and so the relative velocity is determined
 from the steady balance (\ref{TVA}); (b) geostrophic balance holds in
 the remaining momentum equations, which account for the horizontal
 centre of mass velocity (the centre of mass
 executes modified epicycles in the orbital plane);
 and (c) that radial wavelengths of perturbed quantities are long (of
 order $L$ or longer). In so doing, we have reduced the
 order of the system from 6 to 2.

The ingredients for instability,
apparently, are:  a vertical
velocity generated by a vertical pressure gradient (equation [\ref{quadratic system Euler z}]);
accumulation of particles by the associated vertical flux of background particles (equation [\ref{quadratic system COM continuity}]); and finally gas incompressibility (equation [\ref{quadratic system incompressibility}]). This
last equation, though difficult to interpret, is also the location
that drag forces appear explicitly, via \eqref{TVA}.
Lastly, we emphasise again the importance
 of geostrophic balance, which offsets pressure gradients by Coriolis circulation, rather than radial flow. Thus radial pressure gradients do not drive
stabilising radial motions which might alleviate the self-same gradients (but note that vertical pressure gradients remain in full). Therefore, rotation may not make an explicit appearance in these equations but its influence is crucial,
as the next section demonstrates.

\section{Streaming stability in the absence of rotation}
To emphasise the importance of rotation, and geostrophic balance in particular,
let us consider the same problem with neither Coriolis force nor background
shear, that is, in a nonrotating frame, a uniform flow of both dust and fluid,
drifting relative to each other because of a pressure gradient, which is
compensated for by an external gravitational field (and/or an inertial,
non-Coriolis acceleration). 
It is actually more straightforward to handle this problem directly in terms of
the dust and solid perturbations, with none of the approximations used in \S 3. 
 We denote by $\mathbf{v}_{p,g}$ the perturbations of
$\mathbf{V}_{p,g}$, $\delta_p\equiv\delta/f_p$ the logarithmic density
perturbation of the gas, and $h_g\equiv h/f_g$.

 Perturbing equations (\ref{continuity-dust})-(\ref{Euler-gas}) (with $\Omega =
0$), still assuming incompressibility, gives the following system:
\begin{equation}
(-i(\omega-\mathbf{k}\cdot\mathbf{V}_p)+\frac{1}{t_{\rm stop}})\mathbf{v}_p
-\frac{1}{t_{\mathrm{stop}}}\mathbf{v}_g=0
\end{equation}
\begin{eqnarray}
-\frac{\epsilon}{t_{\mathrm{stop}}}\mathbf{v}_p
+(-i(\omega-\mathbf{k}\cdot\mathbf{V}_g)+\frac{\epsilon}{t_{\rm
stop}})\mathbf{v}_g  -\nonumber\\\frac{\epsilon}{t_{\rm
stop}}\delta_p(\mathbf{V}_p-\mathbf{V}_g)+ih_g\mathbf{k}=0
\end{eqnarray}
\begin{equation}
i\mathbf{k}\cdot\mathbf{v}_p-i\omega\delta_p=0
\end{equation}
\begin{equation}
   i\mathbf{k}\cdot\mathbf{v}_g =0
\end{equation}
where $\epsilon \equiv \rho_p/\rho_g$. 
identity matrix.
Setting the determinant of the above 8$\times$8 system to zero, the following
dispersion relation results:
\begin{eqnarray}
\bigg(\omega-\mathbf{k}\cdot\mathbf{V}_p
+\frac{i}{t_{stop}}\bigg)\bigg(\omega^2+(\frac{i}{f_gt_{\mathrm{stop}}}-\mathbf{k}\cdot(\mathbf{V}_p+\mathbf{V}_g))\omega\nonumber\\+(\mathbf{k}\cdot\mathbf{V}_p)(\mathbf{k}\cdot\mathbf{V}_g)-i\frac{\mathbf{k}\cdot\mathbf{V}}{f_gt_{\mathrm{stop}}}\bigg)^2=0
\end{eqnarray}
The first factor corresponds to a mode where the gas' velocity field is
unperturbed, with the gas compensating the disturbance of the dust's with
variation of pressure and density. The quadratic factor (squared) has roots:
\begin{eqnarray}
\omega=\frac{1}{2}\bigg(\mathbf{k}\cdot(\mathbf{V}_p+\mathbf{V}_g)+\frac{i}{f_gt_{stop}}\bigg(-1\pm\nonumber\\\sqrt{(1+i(\epsilon-1)(f_gt_{stop})^2\mathbf{k}\cdot\mathbf{g}_e)^2
-4\epsilon(\mathbf{k}\cdot\mathbf{g}_e(f_gt_{stop})^2)^2}\bigg)\bigg),
\end{eqnarray}
with $\mathbf{g}_e\equiv \nabla P/\rho$. Using the lemma
$\mid\!\mathrm{Re}(\sqrt{(1+ia)^2-b^2})\!\mid \leq 1$ for
$(a,b)\in\mathbb{R}^2$ (YG05), their imaginary part ($s$) is always negative.
For $\mathbf{k}\cdot\mathbf{g}_e(f_gt_{stop})^2 \ll 1$ (as in the terminal velocity approximation limit), these roots may be
expanded as
\begin{equation}
\mathbf{k}\cdot(f_p\mathbf{V}_g+f_g\mathbf{V}_p)-i\epsilon(\mathbf{k}\cdot\mathbf{g}_e)^2(f_gt_{\mathrm{stop}})^3
\label{first root rotationless}
\end{equation}
 and
\begin{equation}
\mathbf{k}\cdot\mathbf{V}-\frac{i}{f_gt_{\mathrm{stop}}}
\end{equation}
 the latter corresponding to a damping where dust and fluid converge in concert
toward equilibrium.

  There is thus no streaming instability in the absence of rotation (as was
also clear from the YG05 numerical calculations).
Rotation, as well as feedback, are both essential for the streaming
instability, hence the difficulty in finding a simple toy model to
explain it.

\section{Toward an interpretation of the streaming instability}
We now develop a framework with which to interpret the streaming instability.
\subsection{Particle concentration in pressure maxima}

  We start with the equation governing the evolution of the dust mass fraction.
Combining equations (\ref{continuity-dust}) and (\ref{continuity-gas}) (we use here the full equations rather than those of the reduced system), we
obtain:
\begin{equation}
\mathcal{D}_p\mathrm{ln}\frac{\rho_p}{\rho}+\frac{\nabla\cdot\left(\rho_g(\mathbf{V}_p-\mathbf{V}_g)\right)}{\rho}=0.
\label{Jacquet continuity}
\end{equation}
It is clear from this equation that dust is advected by $\mathbf{V}_p$
but is also subject to a mass flux associated with the relative
drift (the second term on the left-hand-side of equation (\ref{Jacquet continuity})). We might expect then to observe dust clumps correlated with $\Delta\mathbf{V}$
 (as noted by YG05 in the caption of their figure 6). In the
terminal velocity approximation, equation (\ref{Jacquet continuity}) becomes:
\begin{eqnarray}
\mathcal{D}_p\mathrm{ln}\frac{\rho_p}{\rho}&=&-\frac{1}{\rho}\nabla\cdot\left(\rho_gt_{\mathrm{stop}}\frac{\nabla
P}{\rho}\right)\nonumber\\
&=&\frac{\rho_g}{\rho}\left(\Delta\mathbf{V}\cdot\nabla
\mathrm{ln}\rho-t_{\mathrm{stop}}\frac{\nabla^2 P}{\rho}\right),
\label{Jacquet-theorem-TVA}
\end{eqnarray}
where the latter equality assumes the constancy of $\rho_g t_{\mathrm{stop}}$. 
There are two terms on the right-hand-side: one due to density variation, and one to pressure variation. We shall show below that the latter is responsible for growth, not the former (as was also ruled out by YJ07). 
Equation \eqref{Jacquet-theorem-TVA} may be rewritten in the useful form:
\begin{equation}
\left(\frac{\partial}{\partial
t}+\left(f_p\mathbf{V}_g+f_g\mathbf{V}_p\right)\cdot\nabla\right)\mathrm{ln}\frac{\rho_p}{\rho}=-\frac{\rho_gt_{\rm
stop}}{\rho^2}\nabla^2P
\label{new Jacquet theorem TVA}
\end{equation}
in which we have used incompressibility. Note that we have not yet
linearised the system---this equation holds equally well in the
nonlinear regime---and also that in this subsection we have made no hypothesis on the centre of mass dynamics (e.g. presence of Coriolis forces), since we have only used mass conservation equations and the terminal velocity approximation. Its linearised form, however, with the Fourier
dependence $\exp{[i(k_xx+k_zz-\omega t)]}$, yields
\begin{equation}
\omega=\mathbf{k}\cdot(f_p\mathbf{V}_g+f_g\mathbf{V}_p) +
if_pk^2t_{\mathrm{stop}}\frac{P'}{\rho_p'},
\label{growth-gradP}
\end{equation}
with $P'\equiv\rho h$ and $\rho_p'\equiv\rho\delta$ the perturbation in
pressure and (particle) density, respectively.

What we have in \eqref{new Jacquet theorem TVA} is a form of `advective-reaction'
equation. Dust is advected at a velocity intermediate between
$\mathbf{V}_p$ and $\mathbf{V}_g$, to wit 
\begin{equation} \label{wavy}
f_p\mathbf{V}_{g}+f_g\mathbf{V}_{p}=\mathbf{V}_p+(\mathbf{V}_g-\mathbf{V}) 
\end{equation}
which corresponds to the $U_{\rm sum}$ of YG05, as in the frame used, $\mathbf{k}\cdot\mathbf{V}=0$. 
On the right hand side we have a source term proportional to the pressure
Laplacian. This term will tend to draw dust \emph{towards} pressure maxima in
 the $x-z$ plane 
and \emph{away} from pressure minima. Clearly,
instability is intimately connected with this term, and the
unstable mode works by concentrating dust at pressure maxima. We
discuss both processes in turn.

\subsection{Advection term}
The left-hand side of equation \eqref{new Jacquet theorem TVA} gives rise to the streaming instability 
wave character: a density pattern in $x$ and $z$ will be advected at the 
velocity
\begin{equation}
\mathbf{V}_p + (\mathbf{V}_g-\mathbf{V})
\end{equation}
and in the linear regime this will be entirely radial. The advection 
velocity is the sum of two parts: the particle velocity and a secondary 
contribution issuing from the relative drag (the bracketed term). The first 
term should be familiar, and simply describes the Lagrangian advection of 
particles by the particle velocity itself. The second term is novel. 
Physically, it represents the fact that particle density maxima induce a 
\emph{decrease} in the relative velocity - because the reciprocal drag of 
the two fluids is greater at that location (cf. the first term in equation 
[\ref{Euler-DeltaV}]). This decrease leads to an additional mass flux, which is 
anticorrelated with $\rho_p'$. An outward effective advection is the 
result, which is in addition (and opposition) to that of $\mathbf{V}_p$.
Thus a density peak will tend to be both (a) pushed inward simply by the
primary particle velocity $\mathbf{V}_p$,
and (b) pushed outward by the second drag-induced mass flux that it
has itself excited.
It is easy to show that for $f_g=f_p$ the sum of the two velocities is
precisely zero, i.e. the two drift velocities cancel
perfectly. Conversely, when there are very few particles, 
i.e. $f_p\to 0$, the advection velocity is simply $\mathbf{V}_p$ and
the pattern is carried by the particle fluid alone. This is
because, in this limit, the particles' influence on the gas is so
minor that $\mathbf{V}_g$ is effectively zero. On the other hand,
when there is virtually no gas, $f_g\to 0$, the advection approaches zero, because
$\mathbf{V}_g\to 0$ (as well as $\mathbf{V}_p$) in this limit.

\subsection{Instability term}

\begin{figure}
\resizebox{\hsize}{!}{
\includegraphics{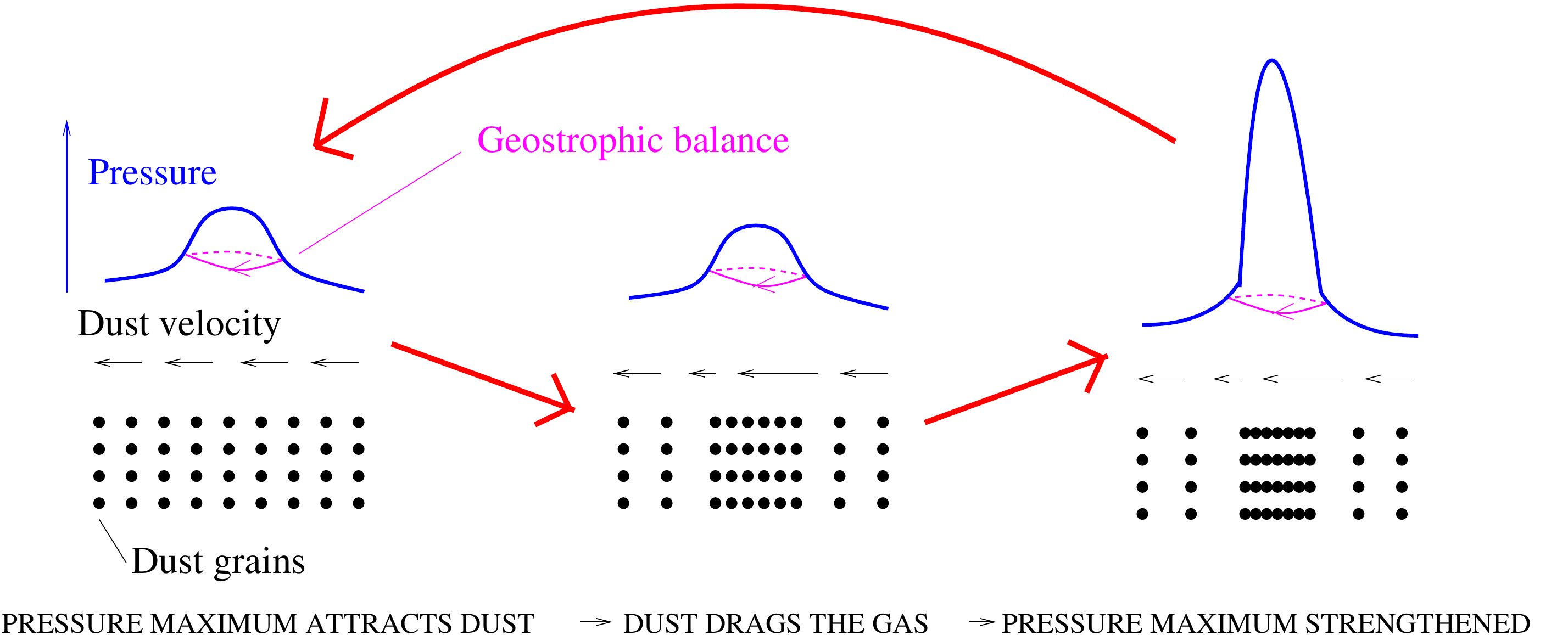}
}
\caption{Cartoon of the proposed interpretation of the streaming instability (seen along the radial axis). Dust is drifting radially inward. The pressure profile is drawn in blue (we ignore the background gradient in the picture). We start with a pressure perturbation maximum. This attracts dust, which in turn drags the gas, creating convergent flows that strengthen the pressure maximum. Note the role of geostrophic balance in maintaining a circulation (schematically shown in magenta) which supports the pressure maximum.}
\label{SI}
\end{figure}

We have already mentioned that instability was related to pressure maxima (or technically, pressure \textit{perturbation} maxima, as there is a background gradient, which however has zero Laplacian
).  This is reminiscent of the concentration mechanism of test particles in long-lived, two-dimensional vortices \citep[e.g.][]{BargeSommeria1995,KlahrBodenheimer2006}, which in a Keplerian flow are anticyclonic.

 In the streaming instability, this concentration is an active process, in the sense that feedback participates in the preservation 
 of the high-pressure, particle-trapping zones. How does the instability grow? 
Schematically, as we have said, pressure maxima attract dust particles, as can be seen from the terminal velocity approximation given by equation \eqref{TVA}. Then, dust drags the gas toward these same maxima (see the gas Euler equation \eqref{Euler-gas}, where the terminal velocity approximation \eqref{TVA} may be injected). This creates convergent flows that \textit{strengthen} these pressure maxima. These in turn can attract more dust particles, and the process thus runs away. This is sketched in Figure \ref{SI}.

  Indeed, inspection of equation \eqref{growth-gradP} shows that growth is correlated with $f_pP'/\rho_p'$. $f_p$ is a measure of the feedback of dust on the gas, while $P'/\rho_p'$ measures how well dust density and pressure maxima correlate with each other, or in other words, the efficiency of the dust-driven pressure loading. Growth is positive if both pressure and density perturbations are not out of phase by more than $90\,^{\circ}$. For example, YJ07's eigenvector ``linA'' has a pressure phase relative to particle density of $43\,^{\circ}$ and a complex frequency $\omega=(-0.3480127+0.4190204i)\Omega$ while that same phase for ``linB'' is $-95\,^{\circ}$ with a complex frequency $\omega=(0.4998786+0.0145764i)\Omega$.\footnote{$s$ is slightly positive, instead of slightly negative, presumably because of finite gas compressibility and/or corrections to the terminal velocity approximation.} For the secular mode in the reduced model, $P'/\rho_p'=h/\delta=(\omega/k_z)^2$ (see equation \eqref{eigenvector}) is real and positive to leading order. 

  While the above identifies the positive feedback loop leading to instability, we have yet to account for two other physical ingredients that are necessary condition to it. Indeed, drag of the dust on the gas is actually insufficient alone to offset the pressure force that tends to accelerate gas \textit{away} from pressure maxima. In the terminal velocity approximation, we have:
\begin{equation}
\frac{\rho_p}{\rho_g}\frac{\mathbf{V}_p-\mathbf{V}_g}{t_{\mathrm{stop}}}-\frac{\nabla P}{\rho_g}=-\frac{\nabla P}{\rho},
\label{drag+gradP}
\end{equation}
The first physical ingredient we wish to mention is rotational dynamics, whose necessity has been shown in \S 3. We have already mentioned in \S 2.3 that a geostrophic balance is established, by which Coriolis forces essentially balance the pressure force. The other physical ingredient is the background solid-to-gas drift (due to the pressure gradient), as noted in \S 2.2. Mathematically, a nonzero drift is required for particle density to affect the dynamics, as, in the reduced system, this can only happen through the ``buoyancy force term'' (proportional to the pressure gradient) in the gas continuity equation \eqref{incompressibility} as mentioned in \S 2.2, which contributes to the advection term discussed in \S 4.2. We suggest that the physical explanation for this is that relative drift between the pressure maximum and the dust allows the former to sweep a (radial) ``headwind'' of particles and be an effective ``trap'' for the dust.

  We note that the processes discussed in the previous paragraph are restricted to the $x$-$y$ plane and emphasise radial drifts and gradients. The fact that the reduced model seems to emphasise vertical gradients---although the leading-order growth rate given by equation \eqref{growth-TVA} is negatively correlated with $k_z$---may be interpreted as resulting from incompressibility, as radial velocity perturbations of the gas must then give rise to vertical perturbations and gradients (see e.g. flow pattern in Fig. 5 of YG05).

\subsection{Streaming instability as collective drag?}
The above demonstration leads us to question the correspondence suggested by
CY10 between the streaming instability and the toy model of
\citet{GoodmanPindor2000}, which was originally designed for a dust layer-scale
instability.
 In this one-dimensional, single-fluid model, the acceleration is the sum of
$g$, a proxy for gravity, pressure and inertial forces, and $-\nu_d(\Sigma)v$ a
density-dependent drag acceleration proportional to the velocity $v$ of the
fluid.
As long as $d\nu_d/d\Sigma \neq 0$, linear analysis shows the system to be
overstable (CY10).

  The fact that this toy model is one-dimensional appears to prevent it from
capturing the specificities of Coriolis forces, unlike other gravitational or
inertial terms. 
 It may also be questionable whether in the linear phase, and in
the local treatment of the instability, the drag can be considered to be
collective in the sense of the above toy model, and what would play the role of
the (single) fluid of this model. Certainly, the latter cannot be the gas+dust
fluid, as drag is internal to it (see equation [\ref{Euler-COM}]). But even if
we take the dust for example, and use the terminal velocity approximation (or
higher-order corrections) as a proxy to eliminate the gas velocity (we cannot
treat it as \textit{imposed}---this is indeed the point of the streaming
instability), the drag acceleration becomes $-\nabla P/\rho$ which makes no
reference to the dust velocity. Same holds for the gas. Finally, as we have noted earlier, the variation of the relative drift with density
is not responsible for growth in this approximation and hence does not conform
to the notion of a collective drag underlying the instability.

\section{Conclusion}

Through successive approximations, we have obtained a reduced system giving
rise to the streaming instability. It would appear that the essence of
the instability can be effectively established given:
\begin{itemize}
\item the terminal velocity approximation
\item that a form of geostrophic balance holds for the planar centre of mass
       velocity
\item that radial wavelengths are sufficiently long.
\end{itemize}
 The leading order (in the stopping time) expansion of the
growth rate is the same as the fourth-order system with the terminal velocity
approximation.

  We interpret the streaming instability as arising from dust pileup in pressure perturbation maxima, with the dust then dragging the gas and hence strengthening the maxima. For this enhancement process to occur despite the pressure force that tends to ``smear'' these maxima out, it is necessary that there be a background solid-gas drift, and also Coriolis forces that help establish a geostrophic balance with the pressure force. The role of Coriolis forces would appear to indicate that one-dimensional toy models do not adequately capture the instability.

\begin{comment}
Though the reduced system does not exhibit rotation explictly, its
importance through geostrophic balance is crucial. As we show in a
system witout rotation:
  Coriolis forces are as important as feedback of the dust
on the gas to understand the streaming instability. This emphasis on
rotation and vertical motion, would appear to indicate that
 one-dimensional toy models do not adequately capture the instability.

The reduced system allows us to offer a physical interpretation of the
instability process. In particular, growth is due to flow
of particles into pressure maxima. While the combination of dust drag
on the gas and geostrophic balance ensures that pressure and density
maxima do not `smear out' radially, vertical motions can help to
reinforce these maxima, at least when the pattern is moving radially
\end{comment}

Larger amplitude vertical flows will (when outside of the linear regime) transport density away from maxima and hence limit the growth of the instability. 
and would compete with other nonlinear effects like particle
clumping ('traffic jams'). Incidentally, another process
 in the nonlinear saturation
that will work against particle clumping is particle pressure, not usually
modelled. The large local densities of nonlinear clumps will lead to enhanced
collision frequencies and consequently to a particle pressure which
will resist further concentration. Attempts to establish the
conditions for self-gravitational collapse of such clumps 
should in principle include this effect.

\bibliographystyle{natbib}
\bibliography{bibliography}

\begin{appendix}

\section{Approximating dust as a pressureless fluid coupled to the
  gas}

It is of interest to investigate in some generality how the description of dust as a pressureless fluid, used in many simulations, arises. As CY10 point
out, the usual criteria for gas molecules 
 do not apply. Intuitively, one expects a fluid description to be viable if the stopping time is short in some sense (YG05, see also YJ07). Here, we verify this intuition using a kinetic approach.
 \citet{Garaudetal2004} addressed this problem for one-dimensional
settling in a static, homogeneous gas, finding that velocity
dispersion would be quickly damped for tightly coupled particles in the Epstein
regime. 

\begin{comment}
  We denote by $\mathbf{F}_p$ the sum of all forces exerted on one particle, moving at velocity $\mathbf{v}_p$, except those forces
caused by particle-particle interactions. 
 Streaming instability calculations ignore such interactions, but
we shall keep them provisionally for more generality. $\mathbf{F}_p$ includes
gas drag on the particle
with $m$ the particle mass.
\end{comment}

We introduce a distribution function $f(\mathbf{r},\mathbf{v}_p)$ of the number of
particles in the ($\mathbf{r}$-$\mathbf{v}_p$) phase space. 
The continuity equation in this space 
may be written as:
\begin{equation}
\frac{\partial f}{\partial t}+\mathbf{v}_p\cdot\frac{\partial
f}{\partial\mathbf{r}} +
\frac{\partial}{\partial\mathbf{v}_p}\cdot\left(f\frac{\mathbf{F}_p}{m_p}\right)=I_{\mathrm{coll}},
\label{myBoltzmann}
\end{equation}
with $\mathbf{F}_p$ the total force exerted on one particle (except particle-particle interactions), $m_p$ the particle mass and $I_{\mathrm{coll}}$ the collision integral. We have used
$\frac{\partial}{\partial\mathbf{r}}\cdot\left(f\mathbf{v_p}\right)=\mathbf{v}_p\cdot\frac{\partial
f}{\partial\mathbf{r}}$ but here, 
$\frac{\partial}{\partial\mathbf{v}_p}\cdot\mathbf{F}_p\neq 0$ because $\mathbf{F}_p$ includes the drag force $-m_p\left(\mathbf{v}_p-\mathbf{V}_g\right)/t_{\mathrm{stop}}$, hence the departure of the
left-hand-side from the standard Boltzmann equation form
\footnote{Thus, technically, what \citet{Garaudetal2004} called an interaction
term, although their approach was collision-free, actually is the corresponding correction to the left-hand side.}.

  Provided collisions conserve the total number and momentum of particles, integration over velocities of equation (\ref{myBoltzmann}) mutliplied by $\mathbf{v_p}$, yields:
\begin{equation}
\mathcal{D}_p\mathbf{V}_p=\frac{1}{m_p}\langle\mathbf{F}_p\rangle-\frac{1}{\rho_p}\nabla\cdot\mathrm{\mathbf{S}},
\label{Euler-dust-general}
\end{equation}
with $\mathbf{V}_p=\langle\mathbf{v}_p\rangle$ the mean particle velocity,
$\mathrm{\mathbf{S}}\equiv
\rho_p\langle\left(\mathbf{v}_p-\mathbf{V}_p\right)\left(\mathbf{v}_p-\mathbf{V}_p\right)\rangle$
the stress tensor. For any particle-wise quantity $x$, we have defined:
\begin{equation}
 \langle x \rangle\equiv\frac{m_p}{\rho_p}\int xf\mathrm{d}^3\mathbf{v}_p.
\end{equation}
Treating the dust as a pressureless fluid amounts to neglecting the stress tensor
in equation (\ref{Euler-dust-general}). To evaluate this, 
it is of interest to derive its evolution equation: 
\begin{eqnarray}
\mathcal{D}_pS^{ij}=m_p\int
I_{\mathrm{coll}}(v_p^i-V_p^i)(v_p^j-V_p^j)\mathrm{d}^3\mathbf{v}_p\nonumber\\-\int
\left(F_p^i(v_p^j-V_p^j)+F_p^j(v_p^i-V_p^i)\right)f\mathrm{d}^3\mathbf{v}_p\nonumber\\-m_p\int
\left(\mathbf{v}_p-\mathbf{V}_p\right)\cdot\frac{\partial
f}{\partial\mathbf{r}}(v_p^i-V_p^i)(v_p^j-V_p^j)\mathrm{d}^3\mathbf{v}_p.
\label{DS/Dt-general}
\end{eqnarray}
The first term on the right-hand-side relates to the redistribution and loss of relative velocities during
collisions, the second to the action of velocity-dependent forces, and the third to
transport of velocity dispersion between neighbouring ``fluid elements''. The
drag contribution to the second term is $-2S^{ij}/t_{\mathrm{stop}}$ (for a velocity-independent $t_{\mathrm{stop}}$). 

  In the absence of particle-particle interaction ($I_{\mathrm{coll}}=0$), integration over space of equation (\ref{DS/Dt-general}) yields:
\begin{eqnarray}
\frac{d}{dt}\langle
S^{ij}\rangle_V=-2\langle\frac{S^{ij}}{t_{\mathrm{stop}}}\rangle_V-\bigg(\epsilon_{\mathrm{ikl}}\Omega^l\langle
S^{kj}\rangle_V+ \langle\frac{\partial V_p^i}{\partial
x^k}S^{kj}\rangle_V\nonumber\\ 
+\epsilon_{\mathrm{jkl}}\Omega^l\langle S^{ki}\rangle_V+ \langle\frac{\partial
V_p^j}{\partial x^k}S^{ki}\rangle_V\bigg)
\label{DS/Dt-avg}
\end{eqnarray}
$\langle...\rangle_V$ denotes a spatial averaging. $\epsilon_{\mathrm{ijk}}$ is
the Levi-Civita tensor and $\mathbf{\Omega}$ is the instantaneous rotation
vector of the reference frame. Damping by gas drag
dominates the evolution of the velocity dispersion if $t_{\mathrm{stop}}$ is
shorter than $\Omega^{-1}$ and the viscous heating rate
$|\nabla\mathbf{V}_p|^{-1}$. This is thus the criterion for validity of the pressureless fluid approximation.

\section{Validity of the terminal velocity approximation}
Perturbing equation (\ref{Euler-DeltaV}) yields:
\begin{eqnarray}
-i\omega\Delta\mathbf{v}-2\Omega\Delta v\mathbf{e}_x+\frac{\Omega}{2}\Delta
u\mathbf{e}_y\nonumber\\-ik_xg_et_{\mathrm{stop}}\left(\mathbf{v}+(f_g-f_p)\Delta\mathbf{v}-f_g\Delta\mathbf{V}\delta\right)\nonumber\\=-\frac{\Delta\mathbf{v}+\delta\Delta\mathbf{V}}{f_gt_{\mathrm{stop}}}+i\frac{h}{f_g}\mathbf{k}
\end{eqnarray}
 (This can also be obtained by injecting equation 31 into equation 27 of YG05).
The terminal velocity approximation amounts to setting the left-hand-side equal
to zero.

  The constraint $\omega f_gt_{\mathrm{stop}}\ll 1$ results from comparing the
first term on the left-hand-side with the first one on the right-hand-side. The
latter dominates the next two terms on the left-hand-side if $\Omega
f_gt_{\mathrm{stop}}\ll 1$, and the fourth and fifth ones if
$f_gk_xg_et_{\mathrm{stop}}^2\ll 1$ (recall that both $f_p$ and $f_g$ are
smaller than 1). To justify the assertion for the fourth term, we combine the
perturbations of (\ref{continuity-COM}) and (\ref{incompressibility})
(equations 28 and 29 of YG05) to obtain:
\begin{equation}
\mathbf{k}\cdot\mathbf{v}=\frac{f_p\mathbf{k}\cdot\Delta\mathbf{v}}{1+\frac{f_gk_xg_et_{\mathrm{stop}}}{\omega}}
\label{k.v}
\end{equation}
and use the expectation that the wave speed is of order the background drift
velocity, as verified by YG05.\footnote{Actually, equation (\ref{k.v}) does not
really constrain $v$, but if we adopt the scaling $v\sim\frac{\Omega}{\omega}u$
from equation (\ref{quartic system Euler y}), the condition for this component
reduces to the already obtained $\Omega f_gt_{\mathrm{stop}}\ll 1$.} The final
term on the left-hand-side is dominated by the second one on the
right-hand-side under the same condition.

  The condition $f_gt_{\rm stop}\ll \Omega^{-1},\omega^{-1}$ is implied anyway
for the validity of the fluid approximation (see Appendix A). Since $\omega <
\Omega$, the condition on $\tau_s$ ($\ll 1/f_g$) is the most stringent one. 
These conditions are somewhat less stringent than those mentioned by YG05
(which are thus sufficient), and more symmetric with respect to dust and gas
(as they converge in concert toward the terminal (relative drift) velocity
because of their mutual interaction).
\end{appendix}

\end{document}